\documentclass[conference]{IEEEtran}
\IEEEoverridecommandlockouts
\usepackage[table]{xcolor}
\usepackage{comment}

\usepackage{scalerel}
\usepackage{tikz}
\usetikzlibrary{svg.path}

\definecolor{orcidlogocol}{HTML}{A6CE39}
\tikzset{
  orcidlogo/.pic={
    \fill[orcidlogocol] svg{M256,128c0,70.7-57.3,128-128,128C57.3,256,0,198.7,0,128C0,57.3,57.3,0,128,0C198.7,0,256,57.3,256,128z};
    \fill[white] svg{M86.3,186.2H70.9V79.1h15.4v48.4V186.2z}
                 svg{M108.9,79.1h41.6c39.6,0,57,28.3,57,53.6c0,27.5-21.5,53.6-56.8,53.6h-41.8V79.1z M124.3,172.4h24.5c34.9,0,42.9-26.5,42.9-39.7c0-21.5-13.7-39.7-43.7-39.7h-23.7V172.4z}
                 svg{M88.7,56.8c0,5.5-4.5,10.1-10.1,10.1c-5.6,0-10.1-4.6-10.1-10.1c0-5.6,4.5-10.1,10.1-10.1C84.2,46.7,88.7,51.3,88.7,56.8z};
  }
}

\newcommand\orcidicon[1]{\href{https://orcid.org/#1}{\mbox{\scalerel*{
\begin{tikzpicture}[yscale=-1,transform shape]
\pic{orcidlogo};
\end{tikzpicture}
}{|}}}}
\usepackage{hyperref} 

\usepackage{amsmath,amssymb,amsfonts}
\usepackage[ruled,vlined]{algorithm2e}
\usepackage{graphicx}
\usepackage{textcomp}
\usepackage{biblatex} 
\usepackage{float}
\usepackage{booktabs}
\usepackage{multirow}

\def\BibTeX{{\rm B\kern-.05em{\sc i\kern-.025em b}\kern-.08em
    T\kern-.1667em\lower.7ex\hbox{E}\kern-.125emX}}
\begin{document}

\title{
A Perception-Driven Approach To Immersive Remote Telerobotics.\\

}

\author{\IEEEauthorblockN{Y. T. Tefera$^{1,2}$\orcidicon{0000-0002-9938-4379},
D. Mazzanti$^{1}$\orcidicon{0000-0002-6749-1540},
S. Anastasi$^{3}$\orcidicon{0000-0003-2163-6684},
D. G. Caldwell$^{1}$\orcidicon{0000-0002-6233-9961},
P. Fiorini$^{2}$\orcidicon{0000-0002-0711-8605}, and
N. Deshpande$^{1}$\orcidicon{0000-0003-4851-5655}
\thanks{This research was conducted in collaboration with the Italian National Worker's Compensation Authority (INAIL).}
}
\IEEEauthorblockA{$^{1}$Istituto Italiano di Tecnologia (IIT), Via Morego 30,
16163 Genova, Italy}
\IEEEauthorblockA{$^{2}$Department of Computer Science, University of Verona, Via Strada Le Grazie 15, 37134 Verona, Italy}
\IEEEauthorblockA{$^{3}$Istituto Nazionale per l'Assicurazione contro gli Infortuni sul Lavoro (INAIL), P.le Pastore 6, 00144 Rome, Italy}
}
\maketitle

\begin{abstract}

Virtual Reality (VR) interfaces are increasingly used as remote visualization media in telerobotics. Remote environments captured through RGB-D cameras and visualized using VR interfaces can enhance operators' situational awareness and sense of presence. However, this approach has strict requirements 
for the speed, throughput, and quality of the visualized 3D data. 
Further, 
telerobotics requires operators to 
focus on their tasks fully, 
requiring high perceptual and cognitive skills. 
This paper shows a work-in-progress framework to address these challenges by taking the human visual system (HVS) as an inspiration. Human eyes use attentional mechanisms to select and draw user engagement to a specific place from the dynamic environment. Inspired by this, the framework implements functionalities to draw users's engagement to a specific place while simultaneously reducing latency and bandwidth requirements. 

\end{abstract}


\section{Introduction}

Immersive remote telerobotics (IRT) 
allows real-time immersive visualization and interaction by the user: perceiving the color and 3D profile of remote environments, while simultaneously interacting with the  robotic agents \cite{stotko2018,mossel2016streaming}.  However, such systems bring several challenges, including high resolution in a 
wide field-of-view (FOV), 
network latency, bandwidth, and perceptual and cognitive constraints. Effective implementations in this field would immeasurably improve the usability and performance for users. 

The human visual system (HVS) has a unique characteristic: it does not need every pixel in the FOV to be rendered at uniformly high quality. It has the highest visual acuity at the center of the FOV, which then drops off towards the periphery. This characteristic can help address the above challenges to some degree. 
In our earlier work in this domain, 
the user's gaze is exploited to divide the acquired 3D data (point-cloud) into concentric conical regions of progressively reducing resolution away from the center of the gaze, termed as \textit{foveated rendering} \cite{tefera2022}. It was shown to improve latency and throughput in the data communication. Some preliminary user trials showed that such 
foveated rendering in VR had minimal impact on the quality of user experience \cite{tefera2022}. Despite these improvements, there are limitations 
when dealing with dynamic 
and unstructured visual information. Foveated rendering may lead to an inability to notice significant visual changes in the peripheral regions of the FOV. This paper builds on the previous work, 
by emphasizing users' attentional mechanisms by adding a real-time scene understanding system and utilizing it to draw the user's attention to desired places in the FOV.



\begin{figure*}[th]
 \centering 
  \includegraphics[width=\textwidth]{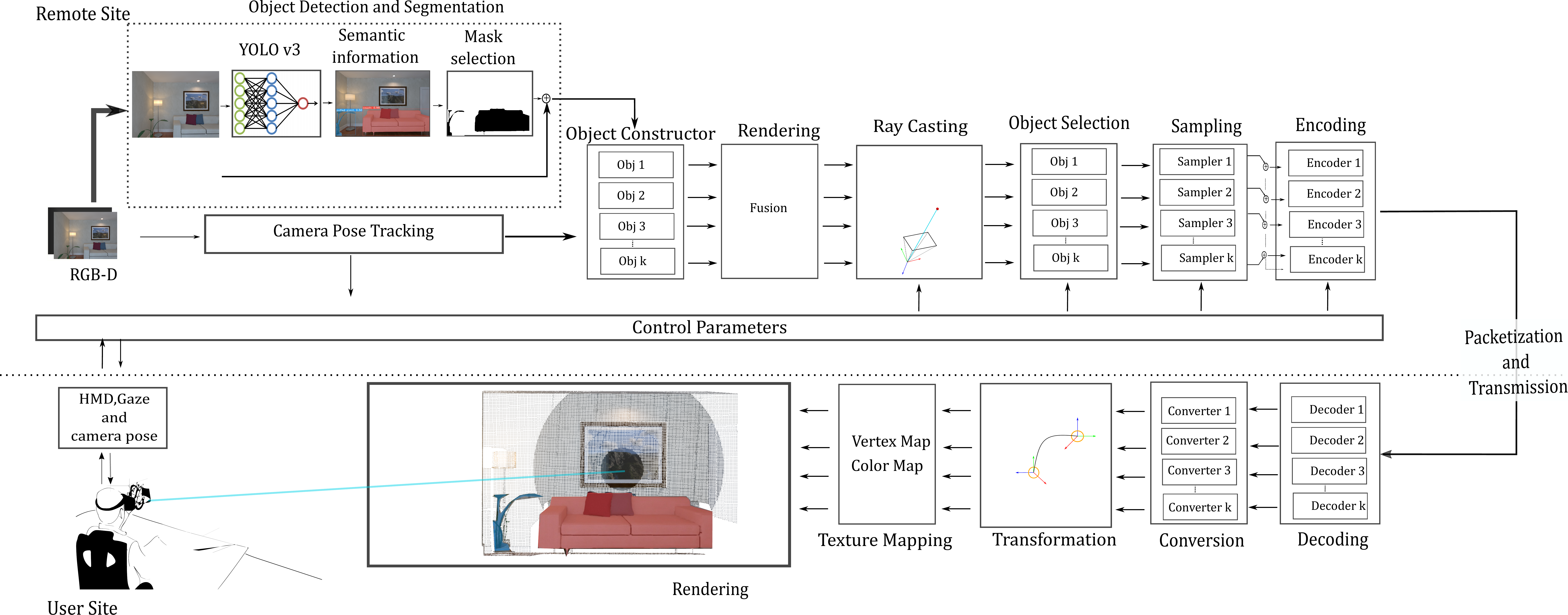}
  \vspace{-6mm}
  \caption{Schema shows an overview of the proposed framework. The remote site performs the 3D data acquisition, semantic scene segmentation, partitioning, foveated sampling, encoding, and parallel streaming. The user site provides the eye-gaze tracking data, decodes the received point-cloud packets, and allows interaction.}\label{fig:system_diagram}
\end{figure*}

\section{SYSTEM OVERVIEW}\label{sec:sys_overview}
The proposed framework shown in  Figure \ref{fig:system_diagram} comprises remote and user site systems that integrate foveated rendering (from our previous work \cite{tefera2022}), with semantic scene understanding. 

\begin{figure}[h]
 \centering 
 \includegraphics[width=0.9\columnwidth]{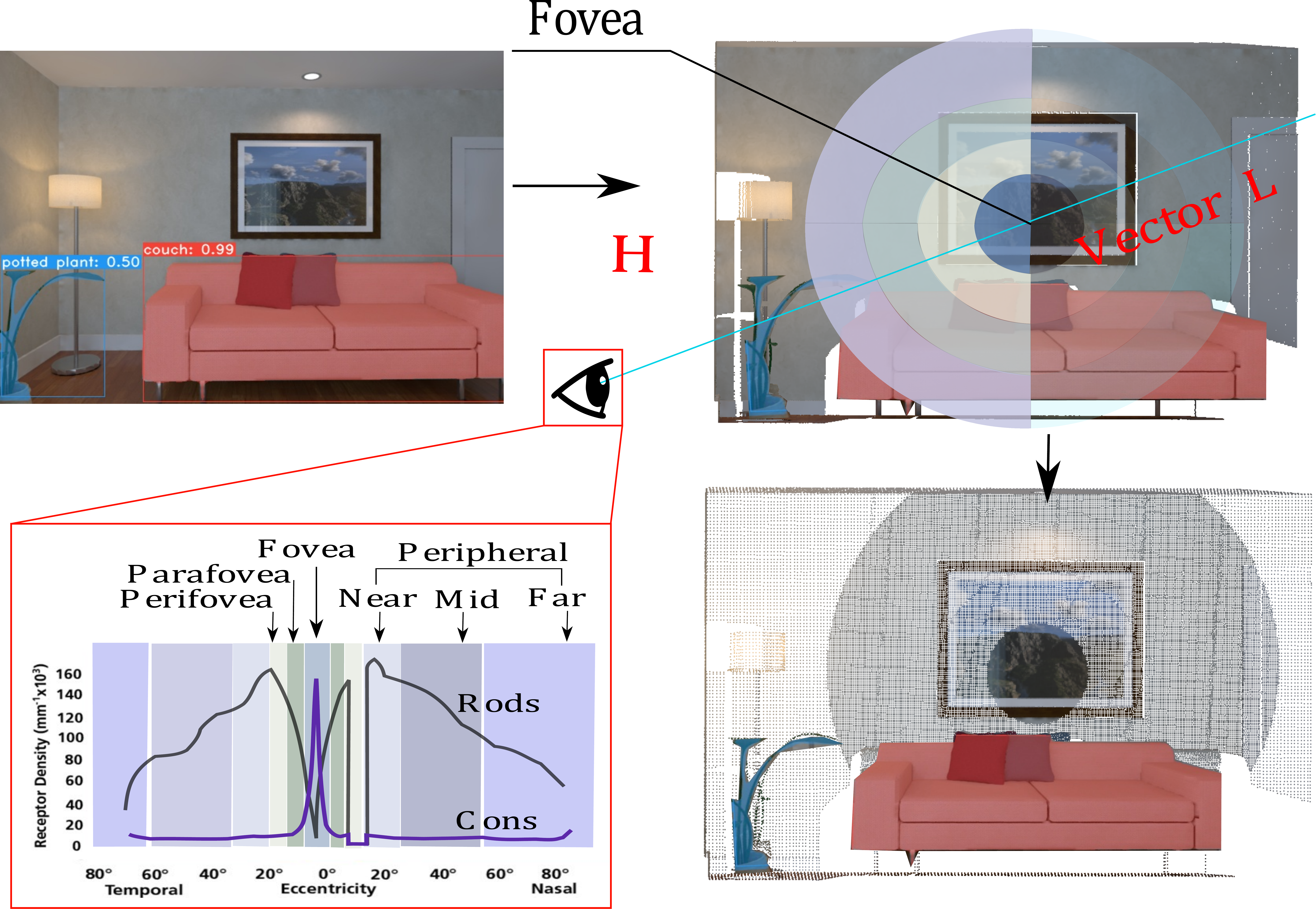}
 \vspace{-2mm}
 \caption{Foveated map partitioning and sampling applied to segmented object maps(bottom-right). Photoreceptor distribution in the retina (bottom-left).}
 \label{fig:semanticComponents}
\end{figure}

The remote site consists of modules implemented for RGB-D data acquisition, semantic scene segmentation, partitioning, foveated sampling, encoding, and parallel streaming. The object detection and segmentation module implement a technique to extract information about the environment and object categories and position information using the state-of-the-art neural network architecture, YOLACT \cite{bolya2020yolact++}, in real-time $(> 30 fps)$. The pipeline takes the input RGB image $\mathbf{C}$ and 
outputs \textbf{N} number of object masks $\mathbf{m}$: $\Omega \to \mathbb{R}$. For all masks $\forall \mathfrak{m}\in{\mathbf{C}}$, it gives a confidence value, $\mathbf{p} \in \{0,100\}$, bounding boxes $ \mathbf{b}:\Omega \to \mathbb{N}^4$, and class IDs $\mathbf{I_t} \in \{0,100\}$, as shown in Figure \ref{fig:semanticComponents}. For all \textbf{N} detected objects, their colour and depth masks are projected as a point-cloud using an unordered list of \textit{surfels}, where each surfel has a position $\mathbf{p} \in \mathbb{R}^3$, a normal $\mathbf{n} \in \mathbb{R}^3$, a colour $\textbf{c}$ $\in$ $\mathbb{R}^3$, a weight $w$ $\in$ $\mathbb{R}$, a radius $r$ $\in$ $\mathbb{R}$, an initialization timestamp $t_0$, and a current timestamp $t$. 

The foveated map partitioning and sampling concept introduced in our work \cite{tefera2022} is extended here. Whereas previously, the whole map was considered as a single entity for partitioning and sampling, here each independent object map $\mathcal{M}_n$ is divided into concentric conical regions. 
Each conical region can then be approximated based on the monotonically decreasing visual acuity in the foveation model, shown in Fig. \ref{fig:semanticComponents} - (bottom-left) \cite{hyona2011foveal}. The $\mathbb{R}^3$ space of each $\mathcal{M}_n$ region is further partitioned into axis-aligned voxels. 
In the previous work \cite{tefera2022}, the foveated sampling would down-sample all the voxels in the peripheral regions uniformly. 
However, this would mean that objects in the peripheral regions would have low quality and would not be noticed by users. Instead, in this paper, with the object-level map, desired objects detected in the periphery can be kept in high resolution, i.e., not down-sampled, thus highlighting them to draw users' attention. A conceptual implementation is shown in Fig. \ref{fig:semanticComponents} - (bottom-right), where the detected couch remains highlighted without down-sampling even if it is in the periphery of the view.

The user site manages the: (1) decoding, conversion, and texture rendering of the streamed 3D data, (2) tracking of the eye-gaze and head-mounted display (HMD) pose, and (3) real-time transfer of gaze and pose information to the remote site. A VR-based interface is designed using the Unreal Engine (UE) on Windows 10, which serves as the \textit{IRT} environment for the user. As shown in Fig. \ref{fig:system_diagram} a parallel streamer, a point-cloud decoder, and a conversion system to transfer the textures to the UE GPU shaders is implemented. A custom point-cloud and data packetization and streaming pipeline was implemented using the Boost ASIO cross-platform C++ library for the communication network.

\section{Experiment Design and Evaluation}

An initial evaluation of the proposed framework is performed using two datasets: a synthetic dataset of a static living room (\textbf{LIV}) and an office room (\textbf{OFF}) \cite{handa:etal:ICRA2014}. Two experimental conditions, \textbf{SEMA-FOV} and \textbf{SEMA}, were created: (i) \textbf{SEMA-FOV} consists of streaming semantically segmented regions, i.e., objects, foveated according to the proposed method; and (ii) \textbf{SEMA} consists of streaming only the semantically segmented regions. These experimental conditions were compared against a reference condition \textbf{REF}, where the original data is streamed without applying the proposed framework. \textit{Data transfer rate} measured using the network data packet analysis tool, Wireshark \cite{sanders2017practical}; and end-to-end \textit{latency} measured for each of the sub-components seen in Fig. \ref{fig:system_diagram} are used to evaluate the proposed framework. Table \ref{table:bandwidth_latency_table} reports the average mean bandwidth and overall latency values for streaming the datasets in each condition: both experimental conditions offer important improvement over the \textbf{REF} condition.

\begin{table}[h]
    \caption{Mean Bandwidth (Mbps) and End-to-End Latency (ms).}
    \label{table:bandwidth_latency_table}
\centering
\begin{tabular}{lcccc}
  \hline
 Dataset & Evaluation & REF  & SEMA-FOV  & SEMA \\ 
  \hline
 \multirow{2}{*}{LIV}& Bandwidth & 2.12  & 1.72 & 0.78  \\ 
                & Latency & 686.83   & 203.92 & 123.15  \\ 

  \hline
\multirow{2}{*}{OFF}& Bandwidth & 1.97  & 1.40 & 0.56  \\ 
                 & Latency & 639.49   & 191.10 & 94.03  \\ 

   \hline
\end{tabular}
\end{table}

\section{Conclusion and Future Work}

The initial evaluation of the FR framework is significant regarding bandwidth and latency improvement. It shows that remotely acquired real-time 3D data can be presented to a user by integrating eye-tracking and semantic segmentation. Future investigations will include a comprehensive user study to measure the impact on quality of experience and user engagement in real-world environments.

\printbibliography
\end{document}